\documentclass{article}
\usepackage{authblk}
\usepackage{graphicx} 
\usepackage{amsmath}
\usepackage{amssymb}
\usepackage{hyperref}
\usepackage{cleveref}
\usepackage{glossaries}
\usepackage{tikz}
\usepackage{svg}
\usetikzlibrary{arrows, arrows.meta, positioning, shapes, backgrounds, fit}
\usepackage{pdflscape}   
\usepackage{booktabs}    
\usepackage{graphicx}    
\usepackage{caption}
\usepackage{booktabs}
\usepackage[a4paper, total={6in, 8in}]{geometry}
\usepackage[utf8]{inputenc}
\usepackage{hyperref}
\usepackage[table]{xcolor}
\usepackage{booktabs}
\usepackage{xspace}
\usepackage{graphicx}
\usepackage{multirow}

\date{\today}

\title{A physics-informed, plug-and-play dose engine for gradient-based radiotherapy treatment planning}
\author[1]{Attila Simkó}
\author[2]{Matthias Kronsteiner}
\author[2]{Simon Glatzer}
\author[1]{Minh Vu}
\author[1]{Josef A. Lundman}
\author[1]{Joakim Jonsson}
\author[1]{Jörgen Olofsson}
\author[1]{Kristina Sandgren}
\author[2]{Wolfgang Lechner}
\author[2]{Dietmar Georg}
\author[3]{Tommy Löfstedt}
\author[1]{Tufve Nyholm}
\author[1]{Anders Garpebring}
\author[2]{Gerd Heilemann}
\affil[1]{Department of Diagnostics and Intervention, Umeå University, Umeå, Sweden}
\affil[2]{Department of Radiation Oncology, Medical University of Vienna, Vienna, Austria}
\affil[3]{Department of Computing Science, Umeå University, Umeå, Sweden}
\date{\today}

\usepackage{biblatex}
\makeglossaries

\begin{document}

\newacronym{AI}{AI}{artificial intelligence}
\newacronym{DL}{DL}{deep learning}
\newacronym{VMAT}{VMAT}{volumetric-modulated arc therapy}
\newacronym[plural=MUs,firstplural=monitor units (MUs)]{MU}{MU}{monitor unit}
\newacronym{MLC}{MLC}{multi-leaf collimator}
\newacronym{TPS}{TPS}{treatment planning system}
\newacronym{MRI}{MRI}{magnetic resonance imaging}
\newacronym{CT}{CT}{computed tomography}
\newacronym{IGRT}{IGRT}{image-guided radiotherapy}
\newacronym{RT}{RT}{radiotherapy}
\newacronym{LMICs}{LMICs}{low and middle income countries}
\newacronym{RL}{RL}{reinforcement learning}
\newacronym{STE}{STE}{straight-through estimator}
\newacronym{WEPL}{WEPL}{water equivalent path length}
\newacronym{OAR}{OAR}{organs-at-risk}
\newacronym{DVH}{DVH}{dose-volume histogram}
\newacronym{HU}{HU}{Hounsfield Unit}
\newacronym{PDRT}{PDRT}{PyDoseRT}

\maketitle

\begin{abstract}
    Radiotherapy treatment planning remains a time-intensive iterative process requiring expert intervention in commercial treatment planning systems (TPS). While machine learning approaches have demonstrated promise, most remain depedent on TPS-based dose calculation or surrogate dose models, preventing direct optimization of deliverable treatment plan parameters. 
    
    We propose \gls{PDRT}, a physics-informed, GPU-accelerated dose engine implemented in PyTorch that computes three-dimensional dose distributions directly from treatment delivery parameters (i.e., MLC leaf positions, jaw positions, gantry angles, and monitor units). The engine preserves gradient information throughout the dose computation pipeline, enabling gradient-based optimization of hardware-constrained treatment plans without the reliance on a commercial TPS. 
    
    \gls{PDRT} was commissioned and validated using water phantom measurements and evaluated on 19 and 162 clinical VMAT prostate cancer plans from two hospitals (with different treatment machines). Central axis depth-dose curves and lateral profiles showed close agreement with measurements (MAE $\leq 1.5\%$). When recalculating clinical plans, \gls{PDRT} achieved high 3D gamma pass rates (mean 96.8~\% for 2~\%/2~mm and 98.9~\% for 3~\%/3~mm, depending on cohort), with Dice coefficients above 0.97 for the $50\%$ isodose volume for both cohorts. DVH analysis demonstrated excellent agreement in mean target and OAR doses (within 0.47~Gy across all structures). A kernel size sensitivity analysis identified a favorable accuracy-performance trade-off, enabling gradient-based optimization with computation times of a few seconds per dose evaluation. All optimized plans converged to clinically acceptable solutions and passed deliverability verification when imported into a commercial TPS. 
    
    This physics-informed framework eliminates TPS dependency for radiotherapy optimization research by enabling gradient-based planning while ensuring that delivery parameters remain in the machine-feasible range. The gradient-enabled dose engine allows exploration of novel optimization strategies and objective functions while maintaining clinical validity. The proposed approach provides a research platform for investigating real-time adaptive radiotherapy concepts, automated planning workflows, and TPS-independent optimization strategies, and democratizing radiotherapy research, by exposing gradient-enabled, hardware-aware, open-source dose computation.
\end{abstract}

\section{Introduction}

Radiotherapy is a central component of modern cancer treatment, with more than half of all patients receiving radiation as part of their therapeutic course. Modern treatment planning involves optimizing thousands of parameters, \textit{i.e.}, \gls{MLC} positions, jaw positions, gantry angles, monitor units, to achieve clinically acceptable dose distributions that satisfy complex and competing dose-volume objectives. This optimization currently still requires an expert making manual iterative adjustments to the optimization parameters in commercial \gls{TPS}. The planning process remains fundamentally manual despite decades of algorithmic advances.

Two major developments expose the limitations of conventional treatment-planning and optimization algorithms. First, the rising global cancer burden, particularly in \gls{LMICs}, will place substantial pressure on radiotherapy services, making it necessary to replace current manual planning steps with automated workflows to maintain capacity and quality \cite{Abdel-Wahab2024, Aggarwal2023}. Second, the widespread adoption of \gls{IGRT} has highlighted the magnitude of daily inter-fraction anatomical variations across a broad range of tumour sites. Although the clinical need for adaptation is evident, practical implementation remains constrained by substantial infrastructural, operational, and resource barriers, even on dedicated IGRT platforms \cite{Qiu2023}. 

\Gls{DL} has transformed medical imaging in radiotherapy, achieving expert-level performance on segmentation \cite{Heilemann2023b}, generating synthetic \gls{CT} images from \gls{MRI} \cite{Fetty2020a, Fetty2020b}, and more. Deep learning models have also been trained for dose distribution predictions from patient anatomy  \cite{Zimmermann2021, gao2025demogenerativeaihelps}. However, a critical gap persists: predicted dose distributions are not treatment plans. Generated dose maps must be imported into a \gls{TPS} for aperture optimization, hence conventional planning stages are still required~\cite{Mekki2025, Chang2025}. Alternatively, parameter prediction methods are \gls{TPS} free, but they learn a direct mapping to delivery parameters without explicitly modeling how parameter variations translate into dose changes, limiting interpretability and gradient-based optimization \cite{Heilemann2023a, Heilemann2025}.

Dose calculation involves complex physics, traditionally implemented in proprietary codebases with limited interfaces for experimentation. While some open-source dose engines exist \cite{Wieser2017, Bhattacharya2025}, they cannot be directly embedded into gradient-based optimization or learning pipelines, where changes in delivery parameters are explicitly mapped to changes in dose.

A consequence is that most \gls{DL}–based approaches have focused on imitating clinically accepted plans generated by conventional \gls{TPS} algorithms. This imitation inherently limits innovation, as it constrains models to replicate the performance of existing planners rather than exceed it. In principle, \gls{DL} could enable exploratory optimization beyond the capabilities of current algorithms; however, only few preliminary studies exist \cite{Mekki2025, Shen2021}. Progress has been hindered in part by the absence of open-source, gradient-enabled dose engines capable of supporting more elaborate optimization strategies.

Another critical limitation concerns computational performance. Online adaptive radiotherapy workflows require dose calculations that are both fast and sufficiently accurate to support real-time plan re-optimization~\cite{Qiu2023}. However, many research directions in treatment planning---particularly gradient-based optimization and \gls{DL}-driven methods---require dose engines that are differentiable and integrated with the optimization process. Commercial \glspl{TPS} do not expose such low-level interfaces, while existing open-source dose calculation frameworks either lack the computational speed needed for online use or do not support differentiation, limiting their applicability for these emerging optimization approaches.

Modern \gls{DL} frameworks (\textit{e.g.}, Pytorch and Tensorflow) enable automatic differentiation: computing gradients of any output with respect to any input through general computational graphs. This capability has revolutionized machine learning, and operations that are implemented with automatic differentiation can be incorporated in a framework that uses gradient-based optimization, such as deep learning models. In radiotherapy, if the dose calculation is differentiable with respect to aperture parameters, we can optimize these parameters directly using gradient descent to achieve the desired dose distributions.

While Liang \textit{et~al.}~\cite{Liang2022} explored Pytorch-based optimization for CyberKnife using a static dose influence matrix, it did not generalize to \gls{VMAT} with full physics simulation and hardware constraint modeling. Similarly, Bhattacharya \textit{et~al.}~\cite{Bhattacharya2025} demonstrated fast GPU-based dose computation for protons, but did not offer gradient access, limiting their utility in optimization and learning tasks.

In this work, we present a modular, gradient-enabled, physics-informed radiotherapy dose engine implemented entirely in Pytorch. The engine decomposes beam transport into modular, GPU-accelerated components---fluence generation from aperture parameters, 3D beam propagation, radiological depth computation, pencil-beam convolution, and final dose reconstruction---each implemented to preserve gradients throughout the computation graph. This design enables optimization of physical plan parameters within the same framework used to define objectives and constraints.

The proposed approach combines the flexibility of modern optimization methods with the physics-based modelling required in radiotherapy planning and dose computation. By guaranteeing gradient backpropagation, modularity, and computational performance, it creates a foundation for new research directions in treatment-planning optimization, including gradient-based aperture refinement, physics-guided machine learning, or automated treatment planning with a trained deep learning model using the dose engine output in a loss function.

Finally, we release the entire framework as open source. This provides the research community with a transparent, extensible platform for investigating next-generation optimization strategies, integrating custom objective functions, and exploring planning paradigms beyond the constraints of conventional treatment-planning systems (\hyperlink{https://github.com/UMU-DDI/PyDoseRT}{https://github.com/UMU-DDI/PyDoseRT}).

\section{Materials and Methods}

\begin{figure}[h!]
\begin{center}
    \centering
    \includegraphics[width=0.94\linewidth]{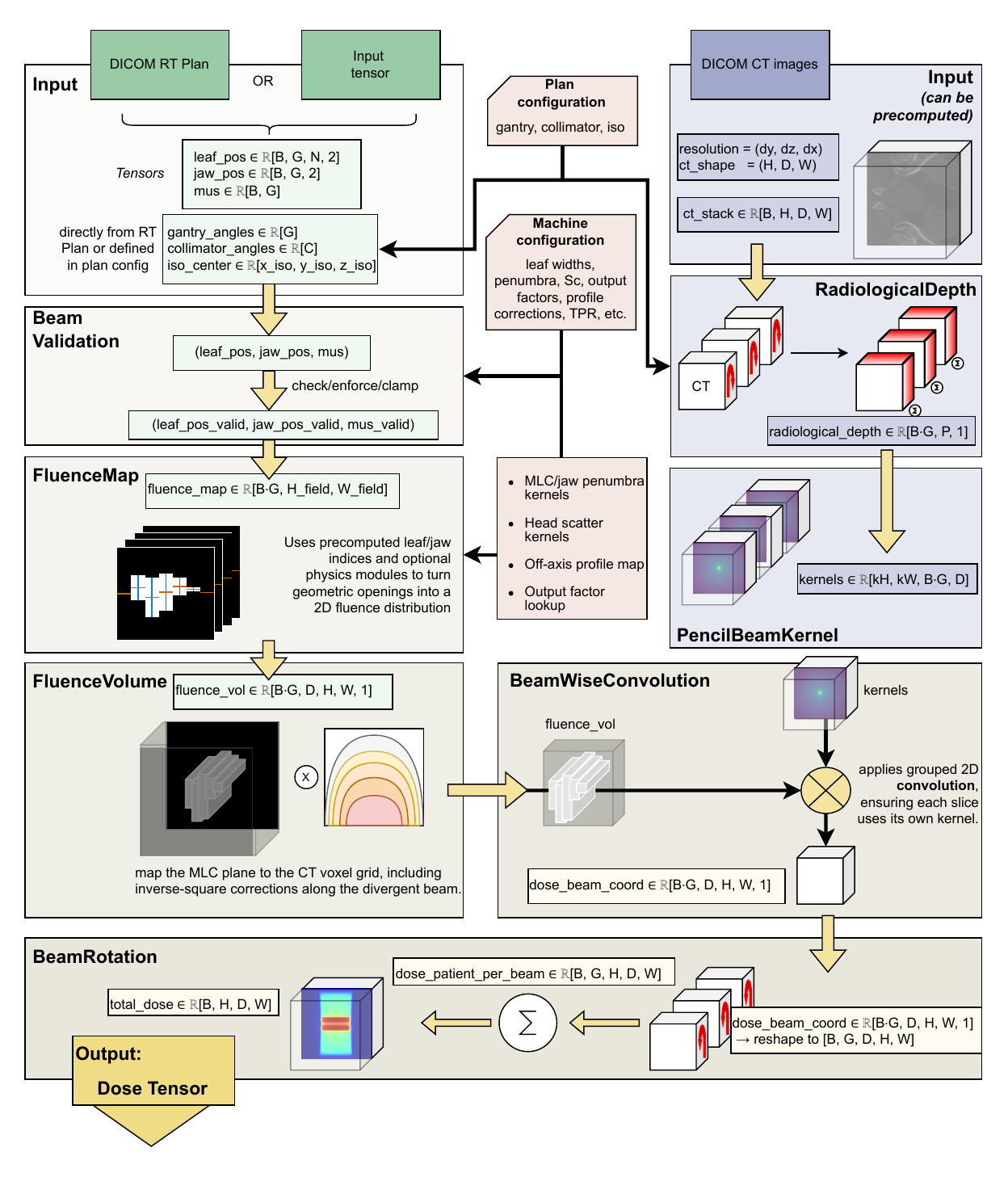}
    \caption{Schematic overview of \gls{PDRT}. Treatment parameters and patient CT are processed through physics-based modules---hardware constrained projection, fluence generation with physics corrections, 3D projection with beam divergence, radiological depth via ray-tracing, depth-dependent kernel convolution, and coordinate transformation---producing final 3D dose distributions while preserving gradients throughout.}
    \label{fig:doseengine_overview}
\end{center}
\end{figure}

\subsection{System Overview}

\textit{\gls{PDRT}} is a gradient-enabled, physics-based pencil-beam dose engine implemented in Pytorch, with its geometric and beam-delivery model tailored to conventional C-arm linear accelerators.
The engine maps treatment delivery parameters---\gls{MLC} leaf and jaw positions and \glspl{MU}---to a 3D dose distribution for a known patient patient \gls{CT} volume and beam geometries. 

The dose calculation pipeline consists of seven modular layers executed in sequence (Fig.~\ref{fig:doseengine_overview}):
\begin{enumerate}
    \item \textbf{BeamValidationLayer}: Projects delivery parameters during optimization into machine-feasible ranges
    \item \textbf{FluenceMapLayer}: Generates 2D fluence maps from aperture parameters with fluence modelling corrections.
    \item \textbf{FluenceVolumeLayer}: Projects fluence maps into 3D volumes accounting for beam divergence
    \item \textbf{RadiologicalDepthLayer}: Computes water-equivalent path lengths via ray-tracing
    \item \textbf{PencilBeamKernel}: Constructs depth-dependent lateral scatter kernels
    \item \textbf{BeamWiseConvolutionalLayer}: Applies kernels via grouped 2D convolutions
    \item \textbf{BeamRotationLayer}: Transforms dose contributions to patient coordinates
\end{enumerate}

A key design principle is the separation of gradient-enabled and gradient-free computations. Layers operating on optimizable parameters (leaf positions, jaw positions, \glspl{MU}) preserve gradients throughout, while operations that depend only on fixed inputs (CT-based radiological depth, kernel generation) are executed outside the gradient graph.

The presented architecture supports standard gradient-based optimizers (\textit{e.g.}, Adam, SGD) for direct optimization of treatment plans, as well as integration with deep learning models where the dose engine serves as a differentiable physics layer within a larger computational graph.

The individual components are described in more detail below.

\subsection{Beam Model and Input Representation}

Each beam is characterized by a parameter vector $\boldsymbol{\theta}_b = (\phi_g, \phi_c, (\ell_i, r_i)_{i=1}^{N_{\text{pairs}}}, (j_x^-, j_x^+), \text{MU}, \mathbf{p}_{iso})$ where $\phi_g$ denotes gantry angle, $\phi_c$ collimator rotation angle, $\ell_i$ and $r_i$ are the left and right MLC leaf positions for the $i$-th leaf pair, $j_x^{\pm}$ represent jaw positions, MU is the prescribed monitor units, and $\mathbf{p}_{\text{iso}} \in \mathbb{R}^3$ specifies the beam isocenter in patient coordinates.

These beams are then organized in a structured list, called a beam sequence, designed for batched operations throughout the dose engine.

For VMAT delivery, the treatment is represented as a sequence of $N_{\text{CP}}$ control points,
\(
\{\boldsymbol{\theta}_b\}_{b=1}^{N_{\text{CP}}},
\)
each corresponding to a discrete gantry angle and machine configuration. Continuous arc delivery is approximated by evaluating dose contributions at these control points, which is consistent with standard VMAT dose calculation approaches. All optimization variables are defined at the level of the control points, ensuring that parameter updates correspond to physically deliverable machine states.

Patient anatomy is represented by a 3D \gls{CT} volume with Hounsfield units converted to relative electron density using clinically commissioned calibration curves.

\subsubsection{Hardware-Constrained Projection (BeamValidationLayer)}

To ensure physically deliverable parameters, unconstrained optimization variables are projected into a machine-feasible subset.

For opposing leaf or jaw pairs with positions $(p_a, p_b)$, we convert to center-width representation,
\begin{equation}
c = \tfrac{1}{2}(p_a+p_b), \qquad w = (p_b-p_a),
\end{equation}
enforce a minimum opening $w \ge w_{\min}$, clamp the aperture center to the lateral travel range $c\in[-F, F]$, where $F$ is the half field width, and reconstruct:
\begin{equation}
\hat{p}_a = c - \tfrac{1}{2}w, \qquad \hat{p}_b = c + \tfrac{1}{2}w.
\end{equation}

Monitor units are constrained to have a non-zero negative value by clamping them to a selected $\epsilon > 0$.For VMAT delivery, $\epsilon$ is derived from machine dynamics:
\begin{equation}
    \epsilon = \frac{\dot{D}_{\min} \cdot \Delta\phi}{\dot{\phi}_{\max}},
\end{equation}
where $\dot{D}_{\min}$ is the minimum dose rate, $\dot{\phi}_{\max}$ is the maximum gantry speed, and $\Delta\phi$ is the angular spacing between control points. This ensures the minimum MU per control point reflects the physical constraint that dose delivery cannot be suspended during arc rotation.

\subsubsection{Fluence Generation (FluenceMapLayer)}

This layer converts delivery parameters to 2D fluence maps by computing fractional aperture overlap with a discrete pixel grid. For each leaf pair $(\ell_i, r_i)$ and pixel at position $d$ with width $\Delta$, the fractional overlap is
\begin{equation}
    f(d; \ell_i, r_i) = \frac{1}{\Delta}\max\left(0, \min(r_i, d + \tfrac{\Delta}{2}) - \max(\ell_i, d - \tfrac{\Delta}{2})\right).
\end{equation}
This piecewise-linear formulation provides stable subgradients for optimization. Jaw collimators define outer field boundaries via analogous masking.

Physics corrections are applied sequentially to the raw aperture fluence:
\begin{itemize}
    \item \textit{MLC transmission}: Residual fluence through closed leaves ($\tau \approx 0.2\%$)
    \item \textit{Source penumbra}: Directional Gaussian convolutions modelling finite source size, with separate kernels for MLC and jaw directions characterized by their FWHM.
    \item \textit{Head scatter}: Field-size-dependent scatter contribution modelled as weighted Gaussian convolutions with amplitude and width parameters
    \item \textit{Off-axis profile}: Radial correction accounting for beam attenuation with distance from central axis
    \item \textit{Residual output factors}: Field-size-dependent corrections interpolated from head scatter measurements.
\end{itemize}

All physics corrections are parameterized by machine-specific coefficients, allowing the dose engine to be commissioned independently for each linear accelerator through standard beam data measurements.

For non-zero collimator angles, the fluence map is rotated in the beam's eye view about the beam central axis to account for the physical rotation of the \gls{MLC} aperture. This in-plane rotation is applied prior to ray-based projection to ensure correct geometric alignment of the fluence pattern with the patient anatomy.

\subsubsection{Fluence Projection into the Patient Volume (FluenceVolumeLayer)}

The 2D fluence maps are forward-projected into the 3D patient volume along diverging rays from the radiation source. For each voxel at distance $d$ from the source, the corresponding fluence sample location accounts for geometric magnification:
\begin{equation}
    \mathbf{u}(d) = \frac{\text{SID}}{d} \cdot \mathbf{x}_{\perp},
\end{equation}
where $\mathbf{x}_{\perp}$ denotes the lateral position in beam coordinates and $SID$ is the source-to-isocenter distance. Sampling uses differentiable bilinear interpolation, and the inverse-square law correction
\begin{equation}
    C(d) = \left(\frac{\text{SID}}{d}\right)^2
\end{equation}
is applied to account for geometric beam divergence.

\subsubsection{Radiological Depth Computation (RadiologicalDepthLayer)}

This layer computes water-equivalent path lengths via ray-tracing along the beam central axis. This approximation assumes lateral homogeneity perpendicular to the beam axis, inherently present in 2D pencil beam kernels. A single ray is traced from the source through the isocenter for each gantry angle, yielding a one-dimensional depth profile:
\begin{equation}
    z(\mathbf{x}) = \int_{\text{source}}^{\mathbf{x}} \rho_{\text{rel}}(s) \, ds,
\end{equation}
where $d$ is the geometric depth along the central axis and $\rho_{\text{rel}}$ denotes relative electron density. The integral is approximated by cumulative summation over uniformly sampled points along each ray, with density values obtained via trilinear interpolation from the CT grid.

This computation is executed under \texttt{torch.no\_grad()}, treating radiological depths as fixed with respect to the CT volume---consistent with the intended use case of optimizing delivery parameters for a fixed patient anatomy.

\subsubsection{Pencil-Beam Kernel Construction (PencilBeamKernelLayer)}

Depth-dependent lateral scatter kernels are generated using an analytical pencil-beam model~\cite{Nyholm2006}. The kernel at each radiological depth $z$ is modeled as a radially symmetric function $K(r; z)$ describing the lateral spread of dose deposition, where $r$ is the radial distance from the beam axis. Kernel shape parameters---governing the amplitude and width of the lateral dose spread---vary with depth according to polynomial expressions whose coefficients are selected based on the beam's tissue-phantom ratio $\text{TPR}_{20,10}$.

Since kernel parameters depend only on fixed commissioning data and precomputed radiological depths, the entire kernel generation occurs outside the gradient computation graph, contributing no overhead to backpropagation.

\subsubsection{Fluence--Kernel Convolution (BeamWiseConvolutionLayer)}

Fluence volumes and depth-dependent kernels are convolved to compute dose deposition. For fluence $\Phi(\mathbf{x}_{\perp}, z)$ at radiological depth $z$, the dose is:
\begin{equation}
    D(\mathbf{x}_{\perp}, z) = \iint \Phi(\mathbf{x}'_{\perp}, z) \, K(\|\mathbf{x}_{\perp} - \mathbf{x}'_{\perp}\|; z) \, d\mathbf{x}'_{\perp}.
\end{equation}
This is implemented using grouped 2D convolutions, where each depth slice is convolved with its corresponding kernel independently. The grouped convolution structure ensures computational efficiency while maintaining per-beam independence.

\subsubsection{Coordinate Transformation (BeamRotationLayer) and Dose Accumulation}

Each beam's dose contribution is rotated back into patient coordinates using a differentiable affine grid transformation with trilinear interpolation. For gantry angle $\phi_g$, the inverse rotation maps patient-frame positions to beam-frame sampling locations.

The final dose distribution is the sum over all control point contributions:
\begin{equation}
    D_{\text{total}}(\mathbf{x}) = \sum_{b=1}^{N_{\text{beams}}} \text{MU}_b \cdot D_b(\mathbf{x}),
\end{equation}
where $\text{MU}_b$ scales each beam's contribution by its prescribed monitor units.

\subsection{Implementation Details and Runtime}

\gls{PDRT} is implemented in Pytorch~2.3 and supports both CPU and GPU computations. Key features include vectorized operations, cached coordinate grids and rotation matrices, optional mixed-precision computation, and sequential beam processing for memory-efficient computations.

\subsection{Commissioning and Parameterization}
\label{sec:commissioning}

Beam model parameters were commissioned using measurements from an Elekta Versa~HD (Medical University of Vienna, Vienna, Austria) and a Varian TrueBeam (Umeå University Hospital, Umeå, Sweden). Commissioned parameters include \gls{HU}-to-density curves, percent depth dose, $\text{TPR}_{20,10}$, source penumbra (MLC/jaw FWHM), head scatter parameters (amplitude, $\sigma$), output factors, off-axis profile corrections, MLC transmission, geometry (leaf widths, travel limits), and mean photon energy calibration factors.

Automatic calibration normalizes output for a set reference \glspl{MU} to 1~Gy at reference conditions (10cm$\times$10cm field, 10~cm depth in water, SSD of 90~cm) by computing dose in a virtual water phantom and adjusting the energy scaling factor.

\subsection{Validation Studies}

\subsubsection{Patient Cohorts and Clinical Plans}

The dose engine was validated on two prostate cancer patient cohorts:
\begin{itemize}
    \item \textbf{Umeå cohort:} 19 prostate cancer patients from the Gold Atlas dataset~\cite{Nyholm2018}. VMAT treatment plans were generated in RayStation (RaySearch AB, Stockholm, Sweden) using 10~MV photon beams and a collapsed cone dose calculation algorithm. All plans were created for delivery on a Varian TrueBeam linac (Varian Medical Systems, Inc., Palo Alto, CA, USA). The prescribed dose to the target was $42.7$Gy delivered in $7$ fractions.
    \item \textbf{Vienna cohort:} 162 prostate cancer patients treated at the Medical University of Vienna. VMAT treatment plans were generated in RayStation using 10~MV photon beams and a collapsed cone dose calculation algorithm. All plans were created for delivery on an Elekta Versa HD linac (ELEKTA AB, Stockholm, Sweden). The prescribed dose to the target was $60$~Gy delviered $20$ fractions. 
\end{itemize}

All clinical plans were exported as DICOM RTPLAN files containing MLC trajectories, jaw positions, gantry angles, and monitor units. All plans met criteria for clinical approval at the respective institutions.

\subsubsection{Water Phantom Validation}

Digital water phantoms (uniform density 1.0~g/cm$^3$) were created to compare \gls{PDRT} against measured dose data from the Varian linac. For each open-field configuration, dose was computed and compared with normalized measurements via central-axis depth-dose profiles, lateral dose profiles at multiple depths (5, 10, 15 cm), and mean absolute error (MAE) values. Sensitivity to kernel resolution was evaluated across multiple kernel sizes.

\subsubsection{Clinical Plan Recalculation}

MLC trajectories, jaw positions, gantry angles, and \gls{MU} were parsed from DICOM RTPLAN files of both patient cohorts and evaluated in \gls{PDRT}. The resulting dose distributions were compared with \gls{TPS} reference doses using global gamma analysis with a 10\% dose cutoff, dose–volume histogram (\gls{DVH}) parameter differences (e.g.\ $D_{95}$, $D_\mathrm{mean}$, $D_\mathrm{max}$), and voxel-wise dose difference metrics. Radiological depth calculations were performed for a single ray parallel to the central beam axis through the isocenter. Consequently, dose accuracy may be reduced in regions where the patient surface has significant obliquity relative to the beam axis, particularly in the build-up region, a known limitation of pencil beam algorithms. To account for this, gamma analysis was performed excluding voxels within 1 cm of the external contour. Further visual comparisons of the dose distributions were performed in Hero Imaging\footnote{\url{https://www.heroimaging.com}} (Hero Imaging AB, Umeå, Sweden).

\subsubsection{Kernel Size Sensitivity}

For five patients from the Umeå cohort, we evaluated dose accuracy and computation time for different kernel sizes. Similarly to the gamma pass rates (2~\%/2~mm), the mean absolute error differences (MAE) were also computed with a 10~\% dose threshold to focus on high dose regions.

\subsubsection{Gradient-Based Optimization}

Starting from closed \gls{MLC} leaves and jaws with zero \glspl{MU}, gradient-based optimization was performed using a simple weighted sum objective:
\begin{equation}
    \mathcal{L} = \lambda_{\text{PTV}} \left\lVert \mathbf{D}_{\text{PTV}} - D_{\text{prescribed}}\right\rVert_{1} + \sum_{s \in \mathcal{S}_{\text{OAR}}} \lambda_s \left\lVert \mathbf{D}_s \right\rVert_{1},
\end{equation}
where $\mathbf{D}_{\text{PTV}}$ and $\mathbf{D}_s$ denote voxel dose vectors within the PTV and structure $s$, respectively, $D_{\text{prescribed}}$ is the prescribed dose, and $\lambda_{\text{PTV}}$, $\lambda_s$ are structure-specific weighting factors. The set of organs at risk $\mathcal{S}_{\text{OAR}}$ included the bladder, rectum, left and right femoral heads, and external contour.

Optimal weighting factors were determined via grid search. The target weight was fixed at $\lambda_{\text{PTV}} = 100$, while OAR weights were selected from $\lambda_s \in \{0, 0.1, 1, 10.0, 100.0\}$. For each weight combination, optimization was run and plan quality was assessed using the Dice coefficient between the 100\% isodose volume (voxels receiving $\geq D_{\text{prescribed}}$) and the PTV contour. The weight combination yielding the highest Dice coefficient was selected. 

The resulting plans were evaluated for clinical quality metrics, computational time, and deliverability by export to DICOM RTPLAN format and verification in a commercial TPS.

\section{Results}

\subsection{Water phantom experiments}

\begin{figure}[ht!]
    \centering
    \includegraphics[width=0.4\linewidth]{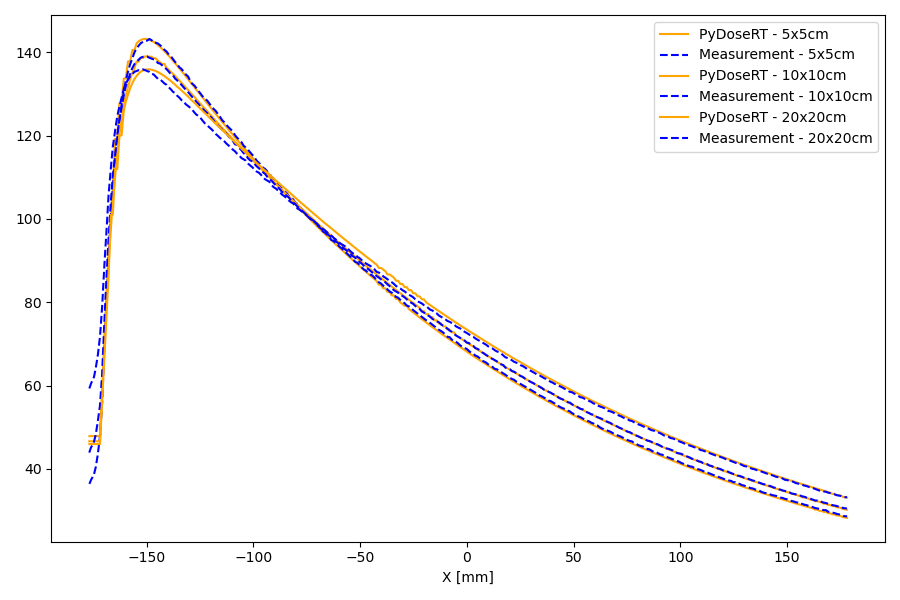}
    \includegraphics[width=0.4\linewidth]{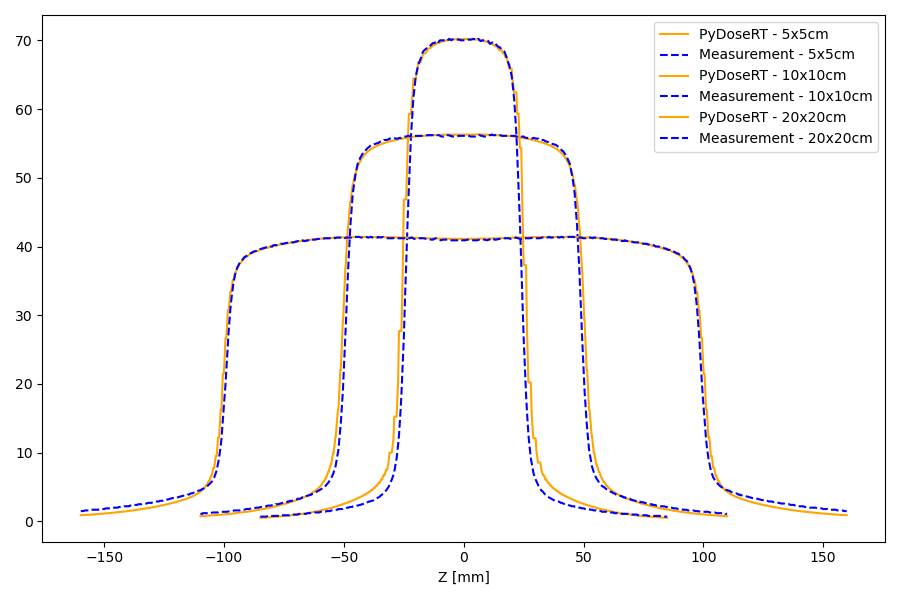}
    \caption{Representative water phantom dose profiles for the Varian TrueBeam linac. Central-axis depth dose curves (left) comparing \gls{PDRT} (solid line) with measurements (dashed). (right) Lateral profiles (right) at 10~cm depth showing central region agreement and penumbra modeling.}
    \label{fig:phantom-profiles}
\end{figure}

A comparison between the profiles is shown in Figure~\ref{fig:phantom-profiles}. Depth-dose curves showed excellent agreement with normalized measurements. The MAE was 0.57~\%, 0.50~\% and 0.91~\% for 5~cm$\times$5~cm, 10~cm$\times$10~cm and 20~cm$\times$20~cm fields, respectively. 

Profile comparisons across multiple field sizes (5~cm$\times$5~cm, 10~cm$\times$10~cm and 20~cm$\times$20~cm) validated the physics corrections for beam divergence, head scatter, and output factors. The MAE was $1.54~\%$, $1.17~\%$, and $0.93~\%$ for the respective fields.

\subsection{Clinical Plan Recalculation}

\subsubsection{Global Dose Agreement}

\begin{figure}[ht!]
    \centering
    \includegraphics[width=\linewidth]{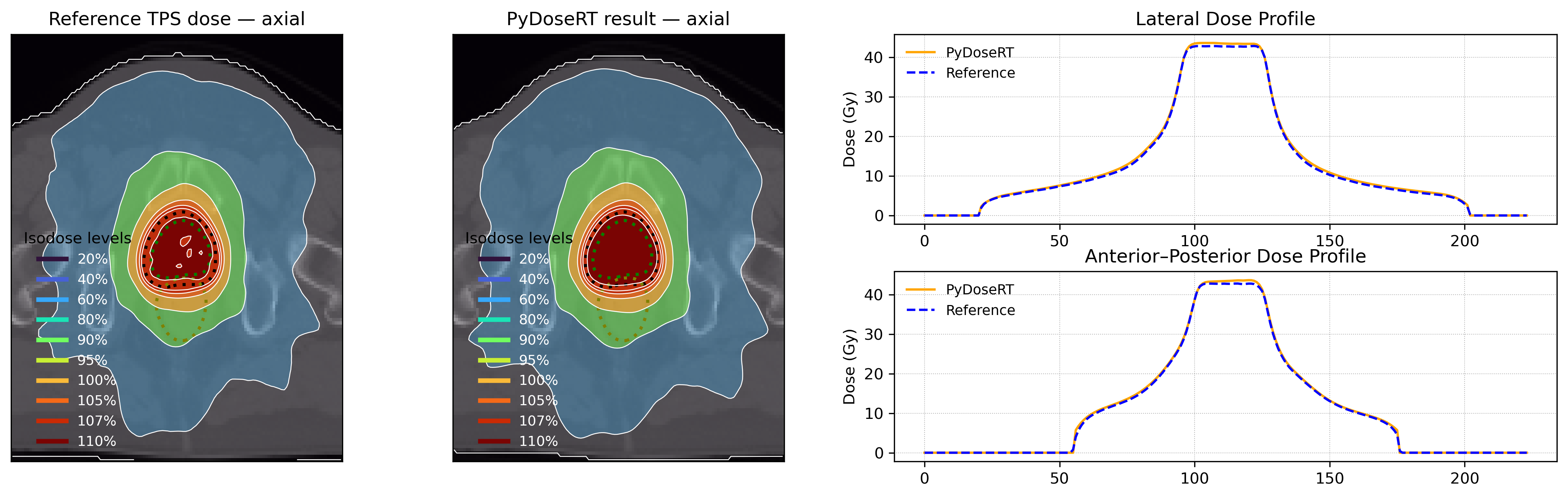}
    \includegraphics[width=\linewidth]{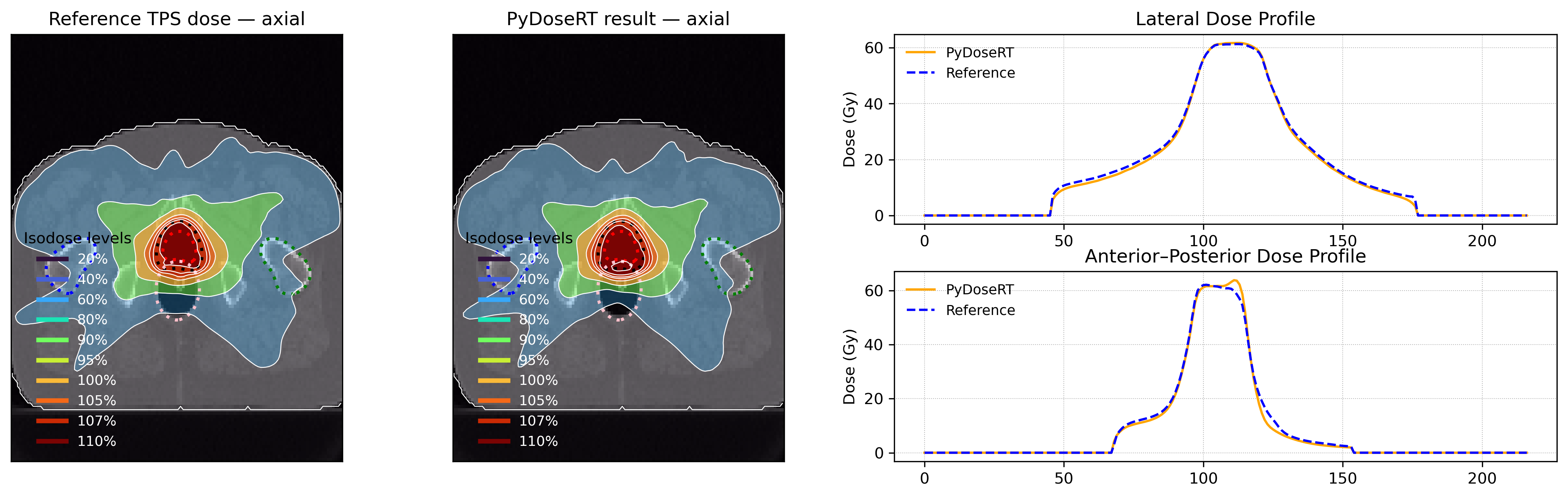}
    \caption{Two examples of dose recalculated with \gls{PDRT} (left) and the \gls{TPS} dose (middle) with two perpendicular line profiles through the isocenter (right) for a patient from the Umeå cohort (top) and from the Vienna cohort (bottom).}
    \label{fig:comarison_examples}
\end{figure}

The dose engine produced accurate dose distributions, compared againts clinical \gls{TPS} doses across both patient cohorts. Table~\ref{tab:global_metrics_comparison} summarizes global agreement metrics. Three-dimensional gamma analysis was performed using a global normalization with a 10~\% dose threshold. For the Umeå cohort, the mean 3D gamma pass rate was 99.99~\% (3~\%/3~mm), so an additional threshold was evaluated, achieving 96.77~\% (2~\%/2~mm). For the Vienna cohort, the mean 3D gamma pase rate was 98.93~\% (3~\%/3~mm). Dice coefficients for the 50~\% and 95~\% isodose were consistently high Table~\ref{tab:global_metrics_comparison}, indicating close agreement of clinically relevant isodose volumes. Dose comparisons are visualized for two patients in Figure~\ref{fig:comarison_examples}.

\begin{table}[h]
\centering
\caption{Global dose agreement metrics between TPS and PDRT-based doses for the Umeå and Vienna patient cohorts (mean [range]).}
\label{tab:global_metrics_comparison}
\begin{tabular}{lrc}
\toprule
\textbf{Metric} & \textbf{Mean [Range]} & \textbf{Unit} \\
\midrule
\multicolumn{3}{l}{\textbf{Umeå cohort}} \\
\midrule
Gamma pass (3\%/3~mm) & 99.996 [99.120--99.999] & \% \\
Gamma pass (2\%/2~mm) & 96.770 [85.965--99.783] & \% \\
Dice 50\% isodose & 0.970 [0.938--0.990] & -- \\
Dice 95\% isodose & 0.976 [0.947--0.989] & -- \\
Mean dose difference & 0.199 [0.131--0.272] & Gy \\
Mean absolute dose difference (MADD) & 0.220 [0.157--0.303] & Gy \\
95th percentile absolute difference & 1.023 [0.762--1.367] & Gy \\
\midrule
\multicolumn{3}{l}{\textbf{Vienna cohort}} \\
\midrule
Gamma pass (3\%/3~mm) & 98.929 [94.528--100.000] & \% \\
Dice 50\% isodose & 0.973 [0.940--0.986] & -- \\
Dice 95\% isodose & 0.944 [0.817--0.983] & -- \\
Mean dose difference & -0.053 [-0.131--0.001] & Gy \\
Mean absolute dose difference (MADD) & 0.065 [0.029--0.136] & Gy \\
95th percentile absolute difference & 0.411 [0.084--0.941] & Gy \\
\bottomrule
\end{tabular}
\end{table}

\subsubsection{Dose-Volume Histogram Comparison}

Tables~\ref{tab:dvh_targets_comparison} and \ref{tab:dvh_oars_comparison}\ summarize \gls{DVH} metric differences between \gls{TPS} and \gls{PDRT} for the Umeå and Vienna cohorts ($\Delta = \mathrm{TPS}-\mathrm{PDRT}$). 

For targets, mean dose agreement was high in both cohorts, with $\Delta D_{\mathrm{mean}}\leq 0.32$~Gy for all CTVs and PTVs. Larger deviations were observed for percentile-based coverage metrics ($D_{98\%}$, $D_{50\%}$, $D_{2\%}$), with mean differences of approximately $1$--$3$~Gy, reflecting the sensitivity of \gls{DVH} percentiles to localized dose differences and steep dose gradients, particularly in the high-dose region. 

For \glspl{OAR}, agreement was strongest for the mean dose metrics, with $\Delta D_{\mathrm{mean}}\leq 0.47$~Gy across all structures and both cohorts. Differences in near-maximum dose metrics ($D_{2\mathrm{cc}}$, $D_{\mathrm{max}}$ were larger, especially for small or high-gradient structures such as the penile bulb and rectum. Volume-based metrics ($V_{50\%}$) showed small absolute differences, typically below $2~\%$, with no systematic bias between \gls{TPS} and \gls{PDRT}. 

\begin{table}[h]
\centering
\caption{Differences in target DVH parameters between TPS and PDRT-based doses for the Umeå and Vienna cohorts (mean [range]). Values represent $\Delta = \text{TPS} - \text{PDRT}$. Note that the prescribed dose to the target for the Umeå cohort was 42.7~Gy and for the Vienna cohort it was 60~Gy.}
\label{tab:dvh_targets_comparison}
\begin{tabular}{lrrrr}
\toprule
\textbf{ROI} & $\Delta D_{98\%}$ [Gy] & $\Delta D_{50\%}$ [Gy] & $\Delta D_{2\%}$ [Gy] & $\Delta D_{\text{mean}}$ [Gy] \\
\midrule
\multicolumn{5}{l}{\textbf{Umeå cohort}} \\
\midrule
CTV & -0.94 [-1.96--0.12] & -0.94 [-2.06--0.13] & -0.94 [-2.15--0.14] & 0.13 [0.01-0.26] \\
PTV & -0.95 [-2.02--0.11] & -0.96 [-2.07--0.11] & -0.94 [-2.18--0.11] & 0.11 [0.00-0.24] \\
\midrule
\multicolumn{5}{l}{\textbf{Vienna cohort}} \\
\midrule
CTV & -1.87 [-8.23-0.45] & -1.99 [-8.30-0.46] & -2.23 [-8.57-0.42] & 0.06 [0.00-0.32] \\
PTV & -2.45 [-8.75-0.47] & -2.60 [-8.84-0.46] & -2.93 [-8.71-0.43] & 0.06 [0.00-0.32] \\
\bottomrule
\end{tabular}
\end{table}

\begin{table}[h]
\centering
\caption{Differences in OAR DVH parameters between TPS and PDRT-based doses for the Umeå and Vienna cohorts (mean [range]). Values represent $\Delta = \text{TPS} - \text{PDRT}$.}
\label{tab:dvh_oars_comparison}
\begin{tabular}{lrrrr}
\toprule
\textbf{ROI} & $\Delta D_{\text{mean}}$ [Gy] & $\Delta D_{2\text{cc}}$ [Gy] & $\Delta D_{\text{max}}$ [Gy] & $\Delta V_{50\%}$ [\%] \\
\midrule
\multicolumn{5}{l}{\textbf{Umeå cohort}} \\
\midrule
Bladder & 0.22 [0.08-0.32] & -0.88 [-2.23--0.12] & 0.13 [0.02-0.32] & -1.37 [-3.24--0.27] \\
Rectum & 0.16 [0.10-0.22] & -0.75 [-3.15--0.14] & 0.11 [0.01-0.45] & -1.04 [-2.32-0.05] \\
Penile Bulb & 0.35 [0.24-0.47] & -1.88 [-3.28--0.26] & 0.27 [0.04-0.47] & -1.30 [-6.67-0.00] \\
Femoral Head L & 0.11 [0.07-0.16] & -0.53 [-0.95--0.17] & 0.08 [0.02-0.14] & -- \\
Femoral Head R & 0.11 [0.08-0.15] & -0.57 [-1.03--0.31] & 0.08 [0.04-0.15] & -- \\
External & 0.08 [0.06-0.12] & -0.93 [-2.15--0.10] & 0.13 [0.02-0.31] & -0.12 [-0.33--0.03] \\
\midrule
\multicolumn{5}{l}{\textbf{Vienna cohort}} \\
\midrule
Bladder & 0.02 [0.00-0.10] & -1.68 [-6.39-2.82] & 0.09 [0.00-0.32] & -0.60 [-4.79-0.89] \\
Rectum & 0.03 [0.00-0.10] & -3.76 [-9.41-0.56] & 0.19 [0.00-0.47] & 1.18 [-3.98-3.82] \\
Femoral Head L & 0.04 [0.00-0.07] & 0.65 [-2.29-2.96] & 0.04 [0.00-0.15] & 0.25 [-0.96-2.74] \\
Femoral Head R & 0.03 [0.00-0.06] & 0.44 [-1.81-2.08] & 0.03 [0.00-0.10] & 0.21 [-0.74-2.60] \\
External & 0.01 [0.00-0.02] & -2.93 [-8.71-0.42] & 0.15 [0.00-0.44] & 0.03 [-0.09-0.18] \\
\bottomrule
\end{tabular}
\end{table}

\subsubsection{Kernel Size Sensitivity}

Gamma pass rates and MAE results are collected in Table~\ref{tab:kernel_sensitivity}. Larger kernels improved gamma pass rates, however the lowest MAE results were achieved with a kernel size of $55$. This was the kernel size selected for the following gradient-based optimization.

\begin{table}[h]
    \centering
    \caption{Kernel size sensitivity analysis: dose accuracy and computation time means and standard deviations for $5$ patients from the Umeå cohort (mean ± std).}
    \label{tab:kernel_sensitivity}
    \begin{tabular}{lccc}
        \toprule
        \textbf{Kernel size [px (cm)]} & \textbf{MAE [cGy]} & \textbf{Gamma Pass [\%]} & \textbf{Time [s]} \\
        \midrule
        225 (45) & $5.01\pm0.93$ & $98.60\pm0.61$ & $14.57\pm2.23$ \\
        125 (25) & $5.00\pm0.92$ & $98.57\pm0.60$ & $7.77\pm0.72$ \\
        75 (15) & $4.49\pm0.86$ & $98.59\pm0.50$ & $4.17\pm0.24$ \\
        55 (11) & $3.61\pm0.76$ & $98.67\pm0.48$ & $3.25\pm0.13$ \\
        25 (5) & $5.45\pm1.09$ & $97.16\pm0.57$ & $2.43\pm0.11$ \\
        15 (3) & $5.75\pm1.13$ & $97.12\pm0.59$ & $2.20\pm0.03$ \\
        \bottomrule
    \end{tabular}
\end{table}

\subsection{Gradient-Based Optimization}

Hyperparameter tuning identified consistent optimal weights across all patients: $\lambda_{\text{bladder}} = 1.0$, $\lambda_{\text{rectum}} = 10.0$, $\lambda_{\text{femur L/R}} = 0.1$, and $\lambda_{\text{external}} = 10.0$.

Starting from closed apertures with zero \glspl{MU}, gradient-based optimization converged to clinically acceptable plans in $100$ iterations. All 19 patients contained CT volumes of different sizes, so the dose grid was also differently sized, however with a kernel size of $55$, optimization took approximately 10 minutes per patient on an NVIDIA A40.

Table~\ref{tab:optimization_results} compares optimized plans with clinical reference plans. PTV coverage and OAR doses were comparable, demonstrating that direct hardware parameter optimization produces deliverable plans meeting clinical criteria. All optimized plans were successfully exported as DICOM RTPLAN files and verified for deliverability in the commercial TPS.

\begin{table}[h]
\centering
\caption{Evaluation of the optimized plans using \gls{PDRT} (mean ± std).}
\label{tab:optimization_results}
\begin{tabular}{lcc}
\toprule
\textbf{Metric} & \textbf{Clinical Plans} & \textbf{Optimized Plans} \\
\midrule
PTV $D_{95\%}$ [Gy] & $41.76\pm0.17$ & $41.47\pm0.24$ \\
PTV $D_{98\%}$ [Gy] & $41.40\pm0.30$ & $40.99\pm0.34$ \\
Rectum $D_{mean}$ [Gy] & $12.78\pm3.55$ & $17.91\pm4.47$ \\
Bladder $D_{mean}$ [Gy] & $11.49\pm5.28$ & $13.38\pm5.50$ \\
Optimization time [min] & $3.3$ & $10.10\pm1.00$ \\
TPS deliverability & 100~\% & 100~\% \\
\bottomrule
\end{tabular}
\end{table}

\begin{figure}
    \centering
    \includegraphics[width=\linewidth]{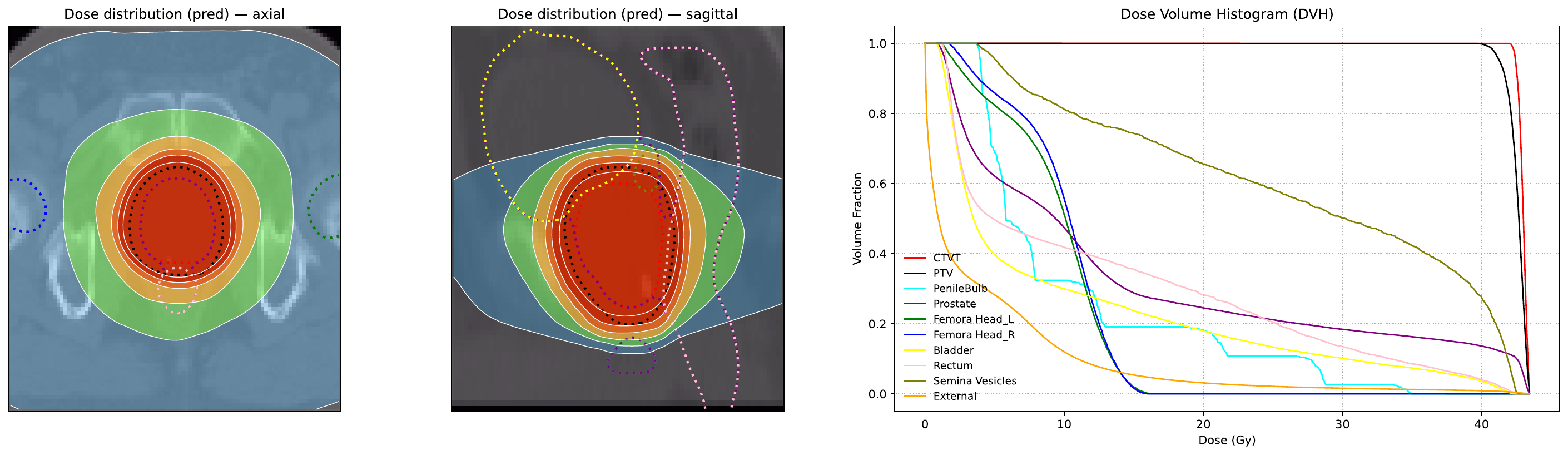}
    \includegraphics[width=\linewidth]{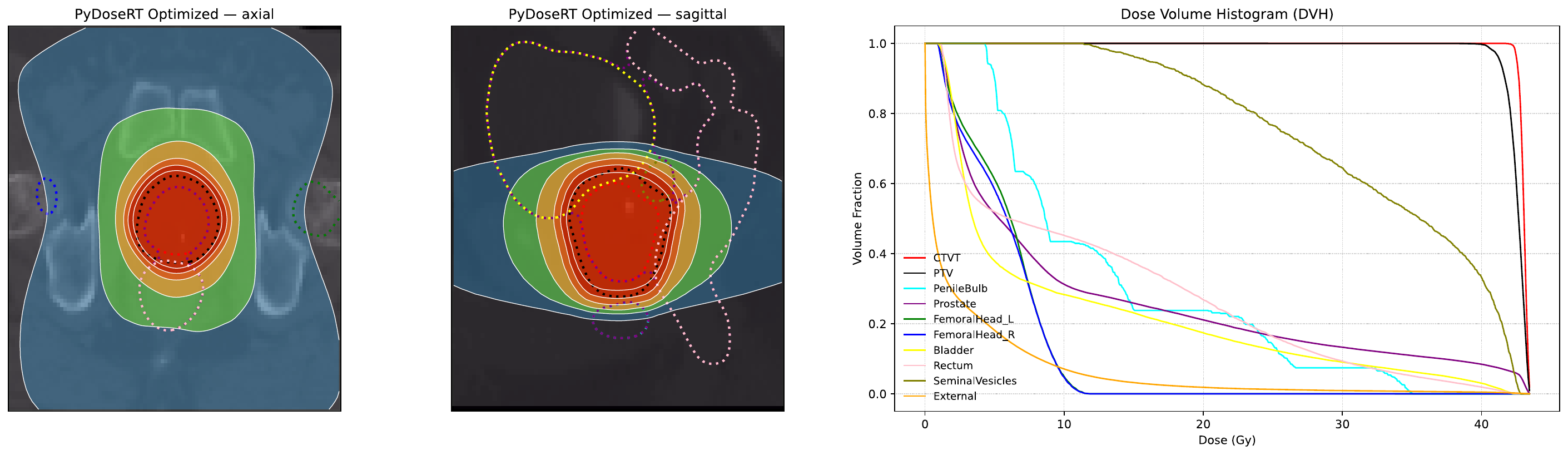}
    \caption{Two example results of gradient-based optimization from the Umeå patient cohort.}
    \label{fig:gradient_optimization_examples}
\end{figure}

\subsection{Computational Performance}

Table~\ref{tab:computational_performance} summarizes computation times for different dose engine operations on an NVIDIA A40 GPU. All computations used a dose grid size of $105\times224\times224$, with the VMAT plans containing 180 control points.

\begin{table}[h]
    \centering
    \caption{Mean computational performance over 100 iterations with a kernel size of $55$. The VMAT plans contain 180 control points computed on a dose grid of size $105\times224\times224$.}
    \label{tab:computational_performance}
    \begin{tabular}{lc}
        \toprule
        \textbf{Operation} & \textbf{Time [s]} \\
        \midrule
        Single beam dose calculation & 0.221 \\
        VMAT calculation step (forward) & 2.051 \\
        VMAT optimization step (forward + backward) & 5.102 \\
        \bottomrule
    \end{tabular}
\end{table}

\section{Discussion}

By implementing the complete dose calculation pipeline in PyTorch, we demonstrate that physics-informed radiotherapy planning---historically locked within proprietary treatment planning systems---can be performed entirely within modern automatic differentiation frameworks. This represents a paradigm shift with implications extending beyond the technical contribution.

Prior work on open-source engines such as matRad \cite{Wieser2017} and \gls{DL}-driven planning frameworks \cite{Mekki2025} have laid the foundation for accessible research tools in radiotherapy, but lack either differentiability, GPU acceleration or seamless integration with modern optimization pipelines. \gls{PDRT} expands this landscape by being the first to offer a transparent, modular, GPU-accelerated dose engine where optimization logic and physical modelling coexist within a single computational graph. Academic groups can develop and validate planning algorithms independently, reserving \gls{TPS} verification for final deliverability confirmation.

Since \gls{PDRT} is implemented entirely in native PyTorch operations, it can be embedded directly into neural network architectures as a differentiable layer. Model outputs representing delivery parameters can be passed through the dose engine, and losses computed on the resulting dose distribution backpropagate gradients to network weights without custom gradient flow implementations. This seamless integration means researchers can use existing \gls{DL} infrastructure---optimizers, schedulers, distributed training---together with \gls{PDRT} as the dose engine behaves as any other PyTorch module.

In contrast to reinforcement learning approaches \cite{Mekki2025, Achlatis2025} that rely on complex policy training and surrogate dose predictors, \gls{PDRT} enables the direct optimization of the delivery parameters based on losses defined on the produced dose, leveraging Pytorch's auto-differentiation to produce clinically valid, hardware-compliant treatment plans.

With dose calculation times of 2--5 seconds per VMAT plan on commodity GPU hardware, iterative optimization becomes feasible within clinically relevant timeframes, in contrast to conventional planning workflows that typically require tens of minutes to hours~\cite{Qiu2023}. This computational efficiency supports emerging use cases such as interactive planning, rapid replanning, and online adaptive \gls{RT}, where all adaptation steps must be completed while the patient remains immobilized on the treatment couch~\cite{Qiu2023, Paganetti2021}. 

To meet these constraints, existing adaptive strategies often rely on plan deformation \cite{Winkel2019}, reoptimization \cite{Archambault2020} or surrogate dose models (i.e., dose prediction) \cite{Babier2022, Zimmermann2021} to achieve acceptable runtimes, which restrict the solution space or decouple optimization from physical dose computation. In contrast, \gls{PDRT} performs repeated, physically grounded dose calculations directly during optimization without requiring an external \gls{TPS}, with few comparable exceptions~\cite{Heilemann2025}. 

The utility of a dose engine for optimization research is determined not only by physical accuracy, but by how efficiently it can be integrated into modern algorithmic workflows. \gls{PDRT} follows this principle by embedding dose calculation directly within an automatic differentiation framework, allowing optimization logic and physical modeling to coexist within a single computational graph. \gls{PDRT} prioritizes computational efficiency and algorithmic accessibility over exhaustive physical modeling. As noted by Bhattacharya \textit{et~al.}~\cite{Bhattacharya2025}, such trade-offs are appropriate for optimization and development workflows, where repeated dose evaluation dominates computational cost, while final clinical verification remains anchored in certified \gls{TPS} implementations.

\subsection{Limitations}

The proposed dose engine is not without its limitations. In several components, design choices favoured simplicity over exhaustive physical modelling, as the modular approach to the proposed framework allows the development and improvement of all components independently, while the overall framework remains robust and thoroughly validated.

One such example is the use of 2D pencil beam kernels. Although more advanced pencil beam kernel methods exist, the chosen implementation was sufficiently accurate for the proof-of-principle experiments presented here and produced plans that were later validated in a \gls{TPS} proving the usefulness of the proposed method. Possible future improvements include 3D kernels, multislab or depth-segmented kernel approaches, which would better account for oblique boundaries.

The dose engine can be used on any input CT, but the area of focus was the prostate, which is a relatively simple use case compared to more advanced treatments, like head-and-neck. Another simplification is the use of fixed gantry angles, instead of delivering dose between control points. For the use case here with $240$ control points, the differences were negligible; however, for many cases, this may be an oversimplification.

During the commisioning process, \gls{PDRT} aimed to implement the most relevant fluence modelling terms, however off-axis softening and electron contamination during the build-up phase are still notably missing.

Finally, although \gls{PDRT} enables gradient-based plan optimization, the primary goal of this work was not to achieve maximal optimization speed, but to establish a flexible framework for developing such algorithms in the future. The demonstrated optimization required approximately $10$ minutes, compared to the corresponding \gls{TPS} optimization times of 3 minutes and 20 seconds. However, further speedups such as initially optimizing with smaller kernel sizes, temporarily disabling time exhaustive fluence modelling steps or employing coarse-to-fine optimization strategies are readily achievable within the framework. In addition, several advanced loss formulations, including DVH-based objectives, are already implemented in \gls{PDRT} but were not explored in this study. These capabilities underscore that the presented results should be interpreted as a baseline demonstration, rather than a performance ceiling.

\subsection{Future directions}

The gradient-enabled framework enables several promising research directions. A natural follow-up for the dose engine is to implement it as a loss function for a neural network that learns treatment delivery parameters from an input \gls{CT} volume, target and \gls{OAR} masks, \textit{e.g.}, as an extension of prior work \cite{Heilemann2023a, Heilemann2025}. Such a model could be trained directly to optimize the dose distribution.

Multi-objective optimization could balance plan quality, delivery efficiency, and robustness simultaneously. Uncertainty quantification through gradient-enabled physics could enable explicit modeling of setup errors and dose calculation uncertainties. Real-time adaptive workflows integrating daily imaging, automatic re-optimization, and deliverability verification could deploy on existing infrastructure.

\section{Conclusions}

We present a gradient-enabled, physics-informed framework for radiotherapy treatment planning implemented in Pytorch. By decomposing dose calculation into modular layers with preserved gradients, the framework enables optimization of machine parameters without commercial TPS dependency.

Comprehensive validation on 181 prostate cancer patients  demonstrated average mean gamma pass rates of $99.99\%$ and $98.93\%$ across two patient cohorts using different treatment machines, and gradient-based optimization with a simple dose-based loss function converged to deliverable plans. The open-source implementation provides a transparent platform for reproducible research.

\section*{Competing Interests}

A.S., J.J., T.N. and A.G. are co-owners of Hero Imaging, the software used for dose quality assessment.

\section{Code \& Data Availability}

All code is available under the MIT license at \hyperlink{https://github.com/UMU-DDI/PyDoseRT}{https://github.com/UMU-DDI/PyDoseRT}. Documentation and tutorials are provided with example data.

The data used for evaluations was the Gold Atlas dataset~\cite{Nyholm2018} for the Umeå cohort, while the Vienna cohort is not publicly available.

\printbibliography

@article{Heilemann2025,
   author = {Gerd Heilemann and Lukas Zimmermann and Tufve Nyholm and Attila Simkó and Joachim Widder and Gregor Goldner and Dietmar Georg and Peter Kuess},
   journal = {Physics and Imaging in Radiation Oncology},
   month = {1},
   publisher = {Elsevier Ireland Ltd},
   title = {Ultra-fast, one-click radiotherapy treatment planning outside a treatment planning system},
   volume = {33},
   year = {2025}
}

@article{Heilemann2023a,
   author = {Gerd Heilemann and Lukas Zimmermann and Raphael Schotola and Wolfgang Lechner and Marco Peer and Joachim Widder and Gregor Goldner and Dietmar Georg and Peter Kuess},
   issue = {8},
   journal = {Medical Physics},
   publisher = {John Wiley and Sons Ltd},
   title = {Generating deliverable DICOM RT treatment plans for prostate VMAT by predicting MLC motion sequences with an encoder-decoder network},
   volume = {50},
   year = {2023}
}

@article{Nyholm2006,
   author = {Tufve Nyholm and Jörgen Olofsson and Anders Ahnesjö and Mikael Karlsson},
   issue = {3},
   journal = {Radiotherapy and Oncology},
   title = {Photon pencil kernel parameterisation based on beam quality index},
   volume = {78},
   year = {2006}
}

@article{Zimmermann2021,
   author = {Lukas Zimmermann and Erik Faustmann and Christian Ramsl and Dietmar Georg and Gerd Heilemann},
   issue = {9},
   journal = {Medical Physics},
   publisher = {John Wiley and Sons Ltd},
   title = {Technical Note: Dose prediction for radiation therapy using feature-based losses and One Cycle Learning},
   volume = {48},
   year = {2021}
}

@article{Qiu2023,
   author = {Zihang Qiu and Sven Olberg and Dick den Hertog and Ali Ajdari and Thomas Bortfeld and Jennifer Pursley},
   issue = {10},
   journal = {Physics in Medicine \& Biology},
   publisher = {Institute of Physics},
   title = {Online adaptive planning methods for intensity-modulated radiotherapy},
   volume = {68},
   url = {https://iopscience.iop.org/article/10.1088/1361-6560/accdb2},
   year = {2023}
}

@article{Heilemann2023b,
   author = {Gerd Heilemann and Martin Buschmann and Wolfgang Lechner and Vincent Dick and Franziska Eckert and Martin Heilmann and Harald Herrmann and Matthias Moll and Johannes Knoth and Stefan Konrad and Inga-Malin Simek and Christopher Thiele and Alexandru Zaharie and Dietmar Georg and Joachim Widder and Petra Trnkova},
   journal = {Physics and Imaging in Radiation Oncology},
   publisher = {Elsevier Ireland Ltd},
   title = {Clinical Implementation and Evaluation of Auto-Segmentation Tools for Multi-Site Contouring in Radiotherapy},
   volume = {28},
   url = {https://linkinghub.elsevier.com/retrieve/pii/S2405631623001069},
   year = {2023}
}

@article{Fetty2020a,
   author = {Lukas Fetty and Mikael Bylund and Peter Kuess and Gerd Heilemann and Tufve Nyholm and Dietmar Georg and Tommy Löfstedt},
   issue = {4},
   journal = {Zeitschrift fur Medizinische Physik},
   publisher = {Elsevier GmbH},
   title = {Latent space manipulation for high-resolution medical image synthesis via the StyleGAN},
   volume = {30},
   year = {2020}
}

@article{Fetty2020b,
   author = {Lukas Fetty and Tommy Löfstedt and Gerd Heilemann and Hugo Furtado and Nicole Nesvacil and Tufve Nyholm and Dietmar Georg and Peter Kuess},
   issue = {10},
   journal = {Physics in Medicine and Biology},
   month = {3},
   pmid = {32235074},
   publisher = {Institute of Physics Publishing},
   title = {Investigating conditional GAN performance with different generator architectures, an ensemble model, and different MR scanners for MR-sCT conversion},
   volume = {65},
   year = {2020}
}

@article{Wieser2017,
   author = {Hans Peter Wieser and Eduardo Cisternas and Niklas Wahl and Silke Ulrich and Alexander Stadler and Henning Mescher and Lucas Raphael Muller and Thomas Klinge and Hubert Gabrys and Lucas Burigo and Andrea Mairani and Swantje Ecker and Benjamin Ackermann and Malte Ellerbrock and Katia Parodi and Oliver Jakel and Mark Bangert},
   issue = {6},
   journal = {Medical Physics},
   month = {6},
   pmid = {28370020},
   publisher = {Wiley Blackwell},
   title = {Development of the open-source dose calculation and optimization toolkit matRad},
   volume = {44},
   year = {2017}
}

@article{Bhattacharya2025,
   author = {Mahasweta Bhattacharya and Calin Reamy and Heng Li and Junghoon Lee and William T. Hrinivich},
   issue = {6},
   journal = {Journal of Applied Clinical Medical Physics},
   month = {6},
   pmid = {40205634},
   publisher = {John Wiley and Sons Ltd},
   title = {A Python package for fast GPU-based proton pencil beam dose calculation},
   volume = {26},
   year = {2025}
}

@article{Abdel-Wahab2024,
   author = {May Abdel-Wahab and C Norman Coleman and Jesper Grau Eriksen and Peter Lee and Ryan Kraus and Ekaterina Harsdorf and Becky Lee and Adam Dicker and Ezra Hahn and Jai Prakash Agarwal and Pataje G S Prasanna and Michael MacManus and Paul Keall and Nina A Mayr and Barbara Alicja Jereczek-Fossa and Francesco Giammarile and In Ah Kim and Ajay Aggarwal and Grant Lewison and Jiade J Lu and Douglas Guedes de Castro and Feng-Ming (Spring) Kong and Haidy Afifi and Hamish Sharp and Verna Vanderpuye and Tajudeen Olasinde and Fadi Atrash and Luc Goethals and Benjamin W Corn},
   issue = {6},
   journal = {The Lancet Oncology},
   month = {6},
   pmid = {38821101},
   publisher = {Elsevier Ltd},
   title = {Addressing challenges in low-income and middle-income countries through novel radiotherapy research opportunities},
   volume = {25},
   url = {https://linkinghub.elsevier.com/retrieve/pii/S147020452400038X},
   year = {2024}
}

@article{Aggarwal2023,
   author = {Ajay Aggarwal and Laurence Edward Court and Peter Hoskin and Isabella Jacques and Mariana Kroiss and Sarbani Laskar and Yolande Lievens and Indranil Mallick and Rozita Abdul Malik and Elizabeth Miles and Issa Mohamad and Claire Murphy and Matthew Nankivell and Jeannette Parkes and Mahesh Parmar and Carol Roach and Hannah Simonds and Julie Torode and Barbara Vanderstraeten and Ruth Langley},
   issue = {12},
   journal = {BMJ Open},
   month = {12},
   pmid = {38149419},
   publisher = {BMJ Publishing Group},
   title = {ARCHERY: a prospective observational study of artificial intelligence-based radiotherapy treatment planning for cervical, head and neck and prostate cancer - study protocol},
   volume = {13},
   year = {2023}
}

@article{Mekki2025,
   author = {Lina Mekki and William T Hrinivich and Junghoon Lee},
   issue = {22},
   journal = {Physics in Medicine \& Biology},
   title = {Dual-arc VMAT machine parameter optimization for localized prostate cancer using deep reinforcement learning},
   volume = {70},
   year = {2025}
}

@article{Shen2021,
   author = {Chenyang Shen and Liyuan Chen and Xun Jia},
   issue = {13},
   journal = {Physics in Medicine and Biology},
   publisher = {IOP Publishing Ltd},
   title = {A hierarchical deep reinforcement learning framework for intelligent automatic treatment planning of prostate cancer intensity modulated radiation therapy},
   volume = {66},
   year = {2021}
}

@article{Nyholm2018,
   author = {Tufve Nyholm and Stina Svensson and Sebastian Andersson and Joakim Jonsson and Maja Sohlin and Christian Gustafsson and Elisabeth Kjellén and Karin Söderström and Per Albertsson and Lennart Blomqvist and Björn Zackrisson and Lars E. Olsson and Adalsteinn Gunnlaugsson},
   issue = {3},
   journal = {Medical Physics},
   title = {MR and CT data with multiobserver delineations of organs in the pelvic area-Part of the Gold Atlas project:},
   volume = {45},
   year = {2018},
}

@article{Liang2022,
   author = {Bin Liang and Ran Wei and Jianghu Zhang and Yongbao Li and Tao Yang and Shouping Xu and Ke Zhang and Wenlong Xia and Bin Guo and Bo Liu and Fugen Zhou and Qiuwen Wu and Jianrong Dai},
   issue = {1},
   journal = {Radiation Oncology},
   publisher = {BioMed Central Ltd},
   title = {Applying pytorch toolkit to plan optimization for circular cone based robotic radiotherapy},
   volume = {17},
   year = {2022}
}

@article{Achlatis2025,
   author = {Stefanos Achlatis and Efstratios Gavves and Jan-Jakob Sonke},
   month = {6},
   title = {Physics-Guided Radiotherapy Treatment Planning with Deep Learning},
   year = {2025}
}

@misc{gao2025demogenerativeaihelps,
      title={Demo: Generative AI helps Radiotherapy Planning with User Preference}, 
      author={Riqiang Gao and Simon Arberet and Martin Kraus and Han Liu and Wilko FAR Verbakel and Dorin Comaniciu and Florin-Cristian Ghesu and Ali Kamen},
      year={2025},
      eprint={2512.08996},
      archivePrefix={arXiv},
      url={https://arxiv.org/abs/2512.08996}, 
}

@article{Paganetti2021,
   author = {Harald Paganetti and Pablo Botas and Gregory C Sharp and Brian Winey},
   issue = {22},
   journal = {Physics in Medicine \& Biology},
   publisher = {IOP Publishing Ltd},
   title = {Adaptive proton therapy},
   volume = {66},
   year = {2021}
}

@article{Winkel2019,
   author = {Dennis Winkel and Gijsbert H. Bol and Petra S. Kroon and Bram van Asselen and Sara S. Hackett and Anita M. Werensteijn-Honingh and Martijn P.W. Intven and Wietse S.C. Eppinga and Rob H.N. Tijssen and Linda G.W. Kerkmeijer and Hans C.J. de Boer and Stella Mook and Gert J. Meijer and Jochem Hes and Mirjam Willemsen-Bosman and Eline N. de Groot-van Breugel and Ina M. Jürgenliemk-Schulz and Bas W. Raaymakers},
   journal = {Clinical and Translational Radiation Oncology},
   publisher = {Elsevier Ireland Ltd},
   title = {Adaptive radiotherapy: The Elekta Unity MR-linac concept},
   volume = {18},
   year = {2019}
}

@article{Babier2022,
   author = {Aaron Babier and Rafid Mahmood and Binghao Zhang and Victor G L Alves and Ana Maria Barragán-Montero and Joel Beaudry and Carlos E Cardenas and Yankui Chang and Zijie Chen and Jaehee Chun and Kelly Diaz and Harold David Eraso and Erik Faustmann and Sibaji Gaj and Skylar Gay and Mary Gronberg and Bingqi Guo and Junjun He and Gerd Heilemann and Sanchit Hira and Yuliang Huang and Fuxin Ji and Dashan Jiang and Jean Carlo Jimenez Giraldo and Hoyeon Lee and Jun Lian and Shuolin Liu and Keng Chi Liu and José Marrugo and Kentaro Miki and Kunio Nakamura and Tucker Netherton and Dan Nguyen and Hamidreza Nourzadeh and Alexander F I Osman and Zhao Peng and Jose Darío Quinto Muñoz and Christian Ramsl and Dong Joo Rhee and Juan David Rodriguez and Hongming Shan and Jeffrey V Siebers and Mumtaz H Soomro and Kay Sun and Andrés Usuga Hoyos and Carlos Valderrama and Rob Verbeek and Enpei Wang and Siri Willems and Qi Wu and Xuanang Xu and Sen Yang and Lulin Yuan and Simeng Zhu and Lukas Zimmermann and Kevin L Moore and Thomas G Purdie and Andrea L McNiven and Timothy C Y Chan},
   issue = {18},
   journal = {Physics in Medicine and Biology},
   publisher = {Institute of Physics},
   title = {OpenKBP-Opt: An international and reproducible evaluation of 76 knowledge-based planning pipelines},
   volume = {67},
   year = {2022}
}

@article{Archambault2020,
   author = {Yves Archambault and Christopher Boylan and Drew Bullock and Tomasz Morgas and Jarkko Peltola and Emmi Ruokokoski and Angelo Genghi and Benjamin Haas and Pauli Suhonen and Stephen Thompson},
   issue = {2},
   journal = {MEDICAL PHYSICS INTERNATIONAL Journal},
   title = {Making on-Line Adaptive Radiotherapy Possible Using Artificial Intelligence and Machine Learning for Efficient Daily Re-Planning},
   volume = {8},
   year = {2020}
}

@article {Chang2025,
	author = {Chang, Ho-hsin and Harms, Joseph and Cardan, Rex Alexander and Fiveash, John B. and Popple, Richard A. and Cardenas, Carlos E.},
	title = {nnDoseNet: Intuitive and Flexible Deep Learning Framework to Train and Evaluate Radiotherapy Dose Prediction Models},
	year = {2025},
	publisher = {Cold Spring Harbor Laboratory Press},
	journal = {medRxiv}
}

\end{document}